\documentclass[a4paper, amsfonts, amssymb, amsmath, reprint,  nofootinbib]{revtex4-1}
\usepackage[english]{babel}
\usepackage[utf8]{inputenc}
\usepackage[colorinlistoftodos, color=green!40, prependcaption]{todonotes}
\usepackage{amsthm}
\usepackage{mathtools}
\usepackage{physics}
\usepackage{xcolor}
\usepackage{graphicx}
\usepackage[left=23mm,right=13mm,top=35mm,columnsep=15pt]{geometry} 
\usepackage{adjustbox}
\usepackage{placeins}
\usepackage[T1]{fontenc}
\usepackage{lipsum}
\usepackage{csquotes}
\usepackage{tikz}
\usepackage{subcaption}

\newcommand{\creation}{\hat{a}^\dagger}
\newcommand{\annihilation}{\hat{a}}
\newcommand{\hamiltonian}
{\hat{\mathcal{H}}}
\newcommand{\raising}{\hat{\sigma}^\dagger}
\newcommand{\lowering}{\hat{\sigma}}

\usetikzlibrary{shadings}
\usetikzlibrary{decorations.pathmorphing}
\usepgflibrary {shadings}
\tikzstyle{charge0}=[top color=green!80!black!50,bottom color=green!80!black,shading angle=20]
\tikzstyle{charge+}=[top color=red!50,bottom color=red!70!black,shading angle=20]
\tikzstyle{charge-}=[top color=blue!50,bottom color=blue!80,shading angle=20]

\usepackage[pdftex, pdftitle={Article}, pdfauthor={Author}]{hyperref} 

\usepackage{xcolor}
\hypersetup{
    colorlinks,
    linkcolor={blue!50!blue},
    citecolor={blue!50!blue},
    urlcolor={blue!50!blue}
}

\usepackage{accents}
\newcommand*{\dt}[1]{%
  \accentset{\mbox{\large\bfseries .}}{#1}}

\begin{document}
\title{Exploring Photon Blockade in Multimode Jaynes-Cummings Models with Two-Photon Dissipation}

\author{Caden McCollum}
\author{Imran M. Mirza}
\email[Correspondence email address: ]{mirzaim@miamioh.edu}
\affiliation{Macklin Quantum Information Sciences, Miami University, Oxford, OH 45056, USA}

\date{\today} 

\begin{abstract}
The photon blockade phenomenon, a promising tool for realizing efficient single-photon sources, is the central focus of our work. We study this phenomenon within the context of the multimode extension of the Jaynes-Cummings model, incorporating two-photon dissipation and external coherent driving. Operating in the weak-driving regime, we confine our analysis to the two-excitation sector of the Hilbert space, initially exploring the single-mode case and then focusing on the corresponding multimode problem. Our study calculates the second-order correlation function (both numerically and analytically) for zero- and nonzero time delays in single- and multimode cases, to pinpoint and validate the conditions that lead to conventional and unconventional photon blockade. Our zero delay findings reveal that photon antibunching is comparable in both cases; however, the multimode case offers a greater degree of control and applicability. Furthermore, for non-zero delay operation, we find that when one of the multiple modes is set at the optimal conventional photon blockade conditions, the behavior of the curve mimics the single-mode problem with an overall slower rate of reaching the $g^{(2)}(\tau)=1$ value. These results highlight the practical implications of our findings for building useful single-photon sources.
\end{abstract}

\keywords{Jaynes-Cummings model, Multimode cavity quantum electrodynamics, Single photon sources, Photon blockade.}

\maketitle


\section{Introduction} \label{sec:intro}
Single-photon sources are increasingly important in the field of quantum information science due to their applications in quantum communication and quantum computing \cite{lounis2005single, o2009photonic, couteau2023applications, knill2001scheme, beveratos2002single, mirza2013single}. However, traditional methods for generating single photons, such as spontaneous parametric down-conversion (SPDC)—in which a nonlinear crystal, such as $\beta$-barium borate, occasionally emits two entangled photons upon the absorption of one energetic photon \cite{couteau2018spontaneous, zhang2021spontaneous, karan2020phase}—can be inefficient. To our knowledge, the highest conversion efficiency achieved in SPDC experiments is approximately $4 \times 10^{-6}$ \cite{bock2016highly}. While higher-efficiency methods for single-photon generation exist, such as the attenuation of coherent light sources \cite{oxborrow2005single, scheel2009single}, these methods may compromise photon antibunching \cite{paul1982photon, lopez2022loss}.

Photon blockade offers a potential solution to these challenges and closely resembles the Coulomb blockade observed in condensed matter systems \cite{bassani2005encyclopedia, grabert2013single}. In a Coulomb blockade, electrons prevent the passage of additional electrons in an electronic system. Similarly, the photon blockade describes a quantum optical system that cannot absorb more photons after reaching a defined excitation threshold \cite{birnbaum2005photon, imamoḡlu1997strongly, zhou2025universal}. This phenomenon can be categorized into three types:
\begin{enumerate}
    \item \textit{Conventional photon blockade} \cite{dayan2008photon, hamsen2017two} occurs due to energy shifts in the atom-field-dressed states and necessitates strong nonlinearity. This form operates in the strong coupling regime of cavity quantum electrodynamics (cQED), where the atom-cavity coupling rate \(g\) must exceed the cavity decay rate \(\kappa\) (i.e., \(g > \kappa\)).
    \item \textit{Unconventional photon blockade} \cite{flayac2017unconventional, snijders2018observation} results from quantum interference, which suppresses the probability of dressed states beyond a certain excitation number. This type of photon blockade can occur in both the strong-coupling regime and the weak-coupling regime (when \(g < \kappa\)).
    \item The third and most recently proposed class is termed \textit{universal photon blockade}. This form relies on optimal conditions in two-photon cQED models, where antibunching can be achieved in all regimes: \(g > \kappa\), \(g < \kappa\), and even \(g \sim \kappa\) \cite{zhou2025universal}.
\end{enumerate}

Previous research on photon blockade in cQED has predominantly employed the Jaynes-Cummings model \cite{jaynes2005comparison, larson2021jaynes}, which characterizes the interaction between a single two-level atom and a single-mode optical cavity. For example, Zhang et al. identified optimal conditions for single-photon generation via both conventional and unconventional photon blockade in two-photon dissipative cQED models, consistently relying on the single-mode framework \cite{zhang2023photon}. Despite its widespread use, the practical realization of optical cavities that support only one mode remains technically challenging \cite{kristensen2014modes}. In addition, recent developments in quantum computing with cQED have demonstrated new quantum phenomena when artificial atoms interact with multiple microwave field modes simultaneously \cite{naik2017random, filipp2011multimode, von2024engineering}.

Motivated by these considerations, this paper presents our novel findings on the photon blockade mechanism in the multimode Jaynes-Cummings model. We discuss how a multimode optical cavity driven by a coherent light source can exhibit photon blockade when all optical modes are coupled to a single two-level atom (qubit or quantum emitter) in the presence of a two-photon dissipation process. Unlike well-studied single-photon dissipation, two-photon dissipation is a nonlinear effect that occurs when two photons, which cannot be absorbed individually, are absorbed simultaneously under overlapping conditions. Although this process is less efficient, two-photon dissipation has been experimentally observed \cite{rumi2000structure, van2002optical}. For our numerical results, we utilize the numerical integration of the Born-Markov master equation \cite{carmichael2007statistical, molmer1993monte}. The master equations for two-photon dissipation have already been reported \cite{garraway1994comparison}.

To investigate photon bunching in characterizing single-photon sources, we calculated the second-order correlation function, \( g^{(2)} \). We begin with the zero-delay correlation function \( g^{(2)}(\tau=0) \) for the single-mode scenario, replicating the results in~\cite{zhang2023photon}. As the first novel aspect of our work, we extend the single-mode analysis to non-zero time delays (when \( \tau \neq 0 \)). Next, as the second and more detailed part of our findings, we examine the second-order correlation function in the multimode regime for \( \tau=0 \) and \( \tau \neq 0 \). In this study, we used the weak drive assumption in the strong-coupling regime of cQED. Our main results are as follows. For the single-mode case, the analytic form of the zero-delay second-order correlation function matches the findings of Zhang et al. For non-zero delay, we numerically calculated and analyzed the second-order correlation function under optimal conditions for both conventional and unconventional photon blockade. In the multi-mode case with tri-modal optical cavities, having multiple modes allows more flexibility in achieving single-photon generation at non-zero delay. In addition, the non-zero delay case exhibits a rapid shift in the photo-emission process toward Poissonian photon statistics.

The remainder of this paper is structured as follows. In the next section, Sec.~\ref{sec:Description}, we provide a theoretical description of our model, detailing the Hamiltonian operator, the second-order correlation function ($g^{(2)}$), and the quantum states for single-mode and multimode problems. In Sec.~\ref{sec:results}, we present our findings for single-mode and three-mode scenarios, discussing both zero-delay and nonzero-delay cases. Finally, in Sec.~\ref{sec:conclusion}, we summarize our key findings.

\section{Theoretical Description} \label{sec:Description}
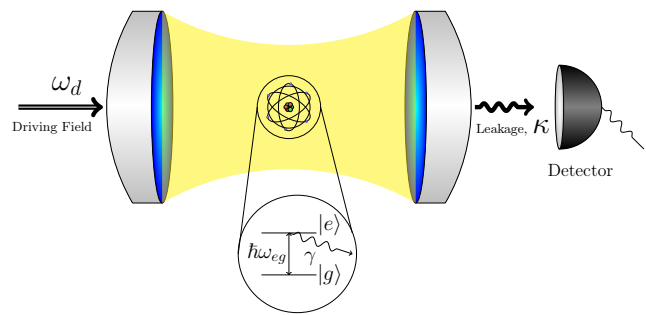
\begin{figure}
    \centering
    \resizebox{\columnwidth}{!}{
    \begin{tikzpicture}[decoration = {snake,   
                    pre length=3pt,post length=7pt,
                    }]  
      \filldraw[top color=lightgray, bottom color=lightgray, middle color=white] (-3.7,-2.286) --(-3,-2.286) -- (-3,2.286)--(-3.7,2.286) arc (150:210:4.572);     
     \filldraw[inner color=cyan, outer color=blue] (-3,0) ellipse (0.1in and 0.9in);

    \filldraw[top color=lightgray, bottom color=lightgray, middle color=white] (3.7,-2.286)--(3,-2.286) -- (3,2.286)--(3.7,2.286) arc(30:-30:4.572);
    \filldraw[inner color=cyan, outer color=blue] (3,0) ellipse (0.1in and 0.9in);
    
\def\d{0.04}
\def\R{0.05}
\def\N{{1.04*(\R+\d)}}
\fill[opacity=0.5,color=yellow] (-3,2.286) arc (240:300:6) -- (3,-2.286)-- (3,-2.286) arc (60:120:6) -- (-3,2.286);
  \def\a{0.2475}
  \def\b{0.5625}
  \def\d{0.06}
  \def\D{0.0675}
  \def\e{0.03} 

  \coordinate (O)  at (  0, 0);
  \coordinate (N1) at (  0:\d);
  \coordinate (N2) at (120:\d);
  \coordinate (N3) at (300:\D);
  \coordinate (P1) at (240:\d);
  \coordinate (P2) at ( 60:\D);
  \coordinate (P3) at (180:\D);
  
  \draw[charge0] (N1) circle (\R);
  \draw[charge0] (N2) circle (\R);
  \draw[charge+] (P1) circle (\R);
  \draw[charge+] (P2) circle (\R);
  \draw[charge0] (N3) circle (\R);
  \draw[charge+] (P3) circle (\R);
  
  \draw[rotate=  0] (0,0) ellipse ({\a} and {\b});
  \draw[rotate=120] (0,0) ellipse ({\a} and {\b});
  \draw[rotate=240] (0,0) ellipse ({\a} and {\b});
  \foreach \i/\ellipse/\theta [evaluate={\x=\a*cos(\theta-\ellipse); \y=\b*sin(\theta-\ellipse);}]
           in {1/0/80,2/0/250,3/120/50,4/120/200,5/240/150,6/240/-50}{
    \fill[charge-,rotate=\ellipse] (\x,\y) circle (\e) coordinate (E\i); 
  }
  

      \draw (-0.65,-4) -- (0.65, -4); 
    \node at (0.97,-4) {\Large $\ket{g}$};
    \draw (-0.65,-3) -- (0.65, -3); 
 \node at (0.97,-2.8) {\Large $\ket{e}$};
 \draw[<->] (0,-4) -- (0,-3);
 \draw[->, decorate] (0,-3) -- (1.5,-3.5);
    \node at (-0.52,-3.5) {\Large $\hbar\omega_{eg}$};
    \node at (0.5,-3.6) {\Large $\gamma$};
\draw (0,0) circle (0.5);
\draw (0,0) circle (0.75);
\draw (0.2,-3.5) circle (1.4);
\draw (0.75,0) -- (1.5,-3);
\draw (-0.75,0) -- (-1.2,-3.5);
\draw[->,double,line width=1.5pt] (-6.4,0) -- (-4.4,0);
\node at (-5.6,-0.5) {{Driving Field}};
\node at (-5.25,0.5) {{\huge $\omega_d$}};
\draw[->,decorate,line width=3.0pt] (4.4,0) -- (5.8,0);
\node at (5.3,-0.5) {Leakage, {\huge $\kappa$}};


\filldraw[top color = black, bottom color = black, middle color = gray] (6.4,1) -- (6.4,-1) arc (-90:90:1);
\filldraw[inner color = white, outer color = lightgray] (6.4,0) ellipse (0.1 and 1);
\draw[decorate] (7.4,0) -- (8.4,-1);
\node at (6.9,-1.5) {\large Detector};
        
    \end{tikzpicture}}
\captionsetup{
format=plain,
margin=1em,
justification=raggedright,
singlelinecheck=false
}
    \caption{(Color online) Schematic diagram of the system considered in this work. A two-level atom trapped inside an optical cavity formed by two reflecting mirrors (with the right mirror being partially transmitting as well). The yellow region in the cavity represents multiple optical modes. The system receives an external coherent drive from left with a frequency of \(\omega_d\), and the photon leakage rate from the right mirror is denoted by \(\kappa\). A perfect detector is oriented toward the right mirror to detect the leaked photons and analyze their correlations.}
    \label{fig:model}
\end{figure}

\subsection{Single-mode cQED case}
\subsubsection{Hamiltonian operator}
In the single-mode problem, we begin with the well-known Jaynes-Cummings model, which effectively describes the interaction between a two-level atom with a transition frequency \(\omega_{eg}\) and a single optical mode (\(N_m = 1\)) trapped within a cavity that has a resonant frequency \(\omega_c\). According to this model, the Hamiltonian operator is expressed in the following form:
\begin{equation}
    \hamiltonian_{JC} = \hbar\omega_{eg} \raising\lowering + \hbar\omega_c \creation\annihilation + \hbar g (\raising\annihilation + \lowering\creation),
\end{equation}
where $\lowering$ is the lowering operator for the two-level atom, $\annihilation$ represents the annihilation operator for the single-mode field and $g$ is a parameter with a real value characterizing the interaction strength between the atom and the field. The non-vanishing commutation and anti-commutation relations are given by: $\left[\hat{a},\hat{a}^\dagger\right]=1$ and $\left\lbrace \hat{\sigma}, \hat{\sigma}^\dagger\right\rbrace = 1$. To incorporate the effect of an external laser drive with frequency \(\omega_d\) and intensity related to the parameter \(\varepsilon\), we introduce the drive Hamiltonian as follows:
\begin{equation}
    \hamiltonian_{d} = \hbar\varepsilon(\creation e^{i\omega_d t} + \annihilation e^{-i\omega_d t}).
\end{equation}
We define the system's Hermitian Hamiltonian, denoted as \(\hamiltonian_{0}\), as the sum of the Jaynes-Cummings Hamiltonian \(\hamiltonian_{JC}\) and the drive Hamiltonian \(\hamiltonian_{d}\). Thus, we have: $\hamiltonian_{0} = \hamiltonian_{JC} + \hamiltonian_{d}$. To account for loss mechanisms — specifically, spontaneous emission occurring at a rate of \(\gamma\) and photon leakage occurring at a rate of \(\kappa\) — we utilize the following Lindblad master equation, which includes two-photon dissipation terms \cite{zhang2023photon}:
\begin{align}
    \frac{d\hat{\rho}_s(t)}{dt} = -\frac{i}{\hbar}\left[\hamiltonian_{0}, \hat{\rho}_s\right] + \frac{\kappa}{2}\mathcal{L}\left[\hat{a}^2\right]\hat{\rho}_s + \frac{\gamma}{2}\mathcal{L}\left[\hat{\sigma}\right]\hat{\rho}_s,
\end{align}
where $\hat{\rho}_s(t)$ is the density operator of the system and $\mathcal{L}[\hat{\mathcal{O}}]\hat{\rho}_s := 2\hat{\mathcal{O}}\hat{\rho}_s\hat{\mathcal{O}}^\dagger - \hat{\mathcal{O}}^\dagger\hat{\mathcal{O}}\hat{\rho}_s - \hat{\rho}_s\hat{\mathcal{O}}^\dagger\hat{\mathcal{O}}$ indicates Lindblad terms \cite{carmichael2007statistical}. Furthermore, employing the quantum jump or trajectory approach necessitates the incorporation of the below-mentioned specific anti-Hermitian terms into the Hamiltonian of the system, which is Hermitian
\begin{align}
    \hamiltonian_{AH} = -\frac{i\hbar\gamma}{2}\raising\lowering -\frac{i\hbar\kappa}{2}\creation{^2}\annihilation^2.
\end{align}
It is worthwhile to note that these terms can be derived by including the relevant collapse operators in the Lindblad master equation, as shown in \cite{garraway1994comparison}. Lastly, by combining the Hermitian and anti-Hermitian Hamiltonians and transforming into a frame that rotates with the drive frequency \(\omega_d\), we obtain the Hamiltonian presented below, which will be used for our analytical calculations
\begin{multline}\label{1modeHam}
    \hamiltonian_{(N_m=1)} = \hbar\Delta_{eg}\raising\lowering + \hbar\Delta_c\creation\annihilation + \hbar g (\raising\annihilation + \lowering\creation) \\ + \hbar\varepsilon(\creation + \annihilation) - \frac{i\hbar\gamma}{2}\raising\lowering - \frac{i\hbar\kappa}{2}\creation{^2}\annihilation^2.
\end{multline}
Here, $\Delta_{eg} = \omega_{eg} - \omega_d$ and $\Delta_c = \omega_c - \omega_d$ are the detunings between the atomic transition frequency and the laser frequency and the cavity frequency and the laser frequency, respectively.

\subsubsection{Second-order correlation function}
As illustrated in Fig.~\ref{fig:model}, we positioned a detector on the right side of the cavity to record the statistics of the photons. Specifically, we calculate the second-order correlation function to evaluate the reliability of our system as a single-photon source. The second-order correlation function is defined as \cite{gerry2023introductory}
\begin{equation}\label{g2def}
    g^{(2)}(\tau) = \frac{\left\langle \hat{E}^-(0) \hat{E}^-(\tau) \hat{E}^+(\tau) \hat{E}^+(0)\right\rangle}{\left\langle \hat{E}^-(0)\hat{E}^+(0) \right\rangle \left\langle \hat{E}^-(\tau)\hat{E}^+(\tau) \right\rangle}.
\end{equation}
In a single-mode cavity of length \( L \), with \( \varepsilon_0 \) representing the electric permittivity of the free space, the operators are defined as follows: $\hat{E}^+ = \sqrt{{\hbar \omega_c}/{\varepsilon_0 L}} \, \annihilation \sin(k x)$ and $\hat{E}^- = (\hat{E}^+)^\dagger$. Using these definitions, the second-order correlation function can be expressed in terms of ladder operators as
\begin{equation}
    g^{(2)}(\tau) = \frac{\left\langle \creation (0)\creation(\tau)\annihilation(\tau)\annihilation(0) \right\rangle}{\left\langle \creation(0)\annihilation(0) \rangle \langle \creation(\tau)\annihilation(\tau)\right\rangle}.
\end{equation}

We will now examine the simplest scenario involving zero delay in the single-mode problem. To facilitate our analytical calculations, we choose to operate in the weak-driving regime, where we can limit the number of excitations in the combined Hilbert space to two. This approach enables us to represent the quantum state of our system as described in \cite{bamba2011origin}
\begin{align}
    \ket{\psi}  =~ & C_{g_0}\ket{g,0} + C_{g_1}\ket{g,1} \nonumber \\
    & +~ C_{e_0}\ket{e,0} + C_{e_1}\ket{e,1} + C_{g_2}\ket{g,2}.
\end{align}
We have introduced a shorthand notation in the last equation, where \(\ket{s, i} \equiv \ket{s} \otimes \ket{i}\) represents the combined state of the atom and the cavity. Here, \(\ket{s}\) denotes the atomic state, and \(\ket{i}\) indicates the cavity state with the number of photons \(i\). Using this quantum state, we can expand the zero-delay second-order correlation function in terms of the probability amplitudes in the following fashion
\begin{equation}\label{g21mode}
    g^{(2)}(0) = \frac{2|C_{g_2}|^2}{\Big[|C_{g_1}|^2 + |C_{e_1}|^2 + 2|C_{g_2}|^2\Big]^2}.
\end{equation}
Applying the time-dependent Schrödinger equation \(i\hbar\partial_t\ket{\psi}=\hamiltonian_{(N_m=1)}\ket{\psi}\), we derive the following set of equations of motion for the probability amplitudes
\begin{subequations}
    \begin{align}
    &i\dt{C}_{g_0} = \varepsilon C_{g_1},&\\
    &i\dt{C}_{g_1} = \varepsilon C_{g_0} + \Delta_c C_{g_1} + g C_{e_0} + \sqrt{2} \varepsilon C_{g_2},&\\
    &i\dt{C}_{e_0} = g C_{g_1} + \Delta_{eg} C_{e_0} + \varepsilon C_{e_1},&\\
    &i\dt{C}_{e_1} = \varepsilon C_{e_0} + (\Delta_{eg} + \Delta_c) C_{e_1} + \sqrt{2} g C_{g_2},&\\
    &i\dt{C}_{g_2} = \sqrt{2} \varepsilon C_{g_1} + \sqrt{2} g C_{e_1} + 2 \Delta_c C_{g_2}.&
    \end{align}
\end{subequations}
Continuing from the calculations outlined in Ref.~\cite{flayac2017unconventional}, we simplify the previously mentioned system of equations while keeping only the terms up to linear order in $\varepsilon$. We also assume that in the steady state, \(C_{g_0} \approx 1\)\footnote{Note that the assumption \(C_{g_0} \approx 1\), used for analytic calculations, implies most of the population stays in the ground state. This leads to a small probability of generating a single photon even if $g^{(2)}(0)$ approaches zero.}. This assumption enables us to reformulate our system of differential equations as follows:
\begin{subequations}\label{ampeq1mode}
    \begin{align}
    &i\dt{C}_{g_1} = \varepsilon + \Delta_c C_{g_1} + g C_{e_0},&\\
    &i\dt{C}_{e_0} = g C_{g_1} + \Delta_{eg} C_{e_0},&\\
    &i\dt{C}_{e_1} = \varepsilon C_{e_0} + (\Delta_{eg} + \Delta_c) C_{e_1} + \sqrt{2} g C_{g_2},&\\
    &i\dt{C}_{g_2} = \sqrt{2} \varepsilon C_{g_1} + \sqrt{2} g C_{e_1} + 2 \Delta_c C_{g_2}.&
    \end{align}
\end{subequations}
Similarly, with these assumptions, Eq.~\eqref{g21mode} takes a simplified form as 
\begin{equation}
    g^{(2)}(0) \approx \frac{2|C_{g_2}|^2}{|C_{g_1}|^4}.
\end{equation}
Finally, we solve equation \eqref{ampeq1mode} under steady state conditions. By inserting the obtained amplitudes, we derive the analytic form of the second-order correlation function in terms of the system parameters which gives us
\begin{equation}\label{singlemodeg2analytic}
   g^{(2)}(0) = \frac{|(g^2 - \Delta_{eg}'\Delta_c)(\Delta_{eg}'{}^2 + \Delta_{eg}'\Delta_c + g^2)|^2}{(|\Delta_{eg}'|^2)^2 |(g^2 - (\Delta_{eg}' + \Delta_c)(\Delta_c - \frac{i\kappa}{2}))|^2}.
\end{equation}
Here, we define \(\Delta_{eg}' = \Delta_{eg} - {i\gamma}/{2}\). As a function of \(\Delta_{eg}\) and \(\Delta_c\), it is evident that the \(g^{(2)}(0)\) function reaches its minimum value when the conditions \(\Delta_c = {g^2}/{\Delta_{eg}}\) and \(\Delta_c = -{g^2}/{\Delta_{eg}} - \Delta_{eg}\) are satisfied. These conditions align with the optimal scenarios reported in previous studies, such as in Ref.~\cite{liang2019antibunching}, which discuss both conventional and unconventional photon blockades.

\subsection{Multi-mode cQED case}
We will now focus on the problem of single-atom coupled optical cavities that can support multiple modes. When there is no direct coupling between the modes, these modes can interact indirectly through a single atom. Therefore, we will extend the single-mode Hamiltonian (see Eq.~\eqref{1modeHam}) to accommodate a multimode scenario, where the number of modes, denoted as \(N_m\), is equal to \(n\). Thus, we write
\begin{align}
    &\hamiltonian_{(N_m = n)} = \hbar\left(\Delta_{eg}-\frac{i\hbar\gamma}{2}\right)\raising\lowering + \sum \limits^n_{i=1} \Bigg[\hbar\Delta_{c_{i}}\creation_i\annihilation_i +\nonumber\\
    &  \hbar g_i\left(\raising\annihilation_i + \lowering\creation_i\right) + \hbar\varepsilon\left(\creation_i + \annihilation_i\right) - \frac{i\hbar\kappa}{2}\creation_i{^2}\annihilation_i^2\Bigg].
\end{align}
For the \(i\)th cavity mode, we define \(\Delta_{ci} = \omega_{i} - \omega_d\), where \(\omega_{i}\) is the frequency of the mode, \(\annihilation_i\) is the annihilation operator and \(g_i\) represents the interaction strength between the atom and the $i$th optical mode. For simplicity, we assume that all modes experience two-photon dissipation at the same rate \(\kappa\) and that each mode is driven by the same laser drive strength \(\varepsilon\). The non-vanishing commutation relations between the mode ladder operators are given by \([\hat{a}_i, \hat{a}^\dagger_j] = \delta_{ij}\). 

For the n-mode problem, the net $\hat{E}^+$ operator at position $x$ within an optical cavity takes a generalized form as 
\begin{equation}\label{Emulti}
    \hat{\vb{E}}^+ = \vb{e}\sum\limits^n_{i=1}\sqrt{\frac{\hbar\omega_{i}}{\varepsilon_0 L}}\annihilation_i \sin{(k_i x)}.
\end{equation}
In this context, \( k_i \) represents the wavenumber associated with mode \( i \). According to the linear dispersion assumption, the relationship between the wavenumber \( k_i \) and the mode frequency \( \omega_i \) is given by the equation \( \omega_{i} = c k_i \), where \( c \) is the speed of light. The vector \( \mathbf{e} \) indicates the polarization direction of the modes; however, for simplicity, we will ignore this aspect moving forward. Additionally, it is important to recognize that, in the multimode case, the specific variation of the field present in both the last equation and the definition of the atom-mode coupling strength cannot be overlooked. To account for this significant feature, we express the coupling strength between the atom and any \( i \)th mode (denoted as \( g_i \)) in relation to the interaction strength between the atom and the fundamental mode \( g_1 \) as follows:
\begin{equation}\label{Eq10purane}
    g_i = \sqrt{i} \frac{\sin{(k_i x)}}{\sin{(k_1 x)}}g_1,~\forall~i\geq 2.
\end{equation}

\begin{figure}
    \centering
    \captionsetup{
format=plain,
margin=1em,
justification=raggedright,
singlelinecheck=false
}
    \includegraphics[width=\linewidth]{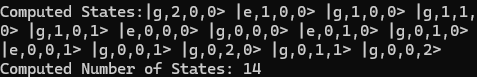}
    \caption{(Color online) Algorithmically computed basis states for a cavity QED system with a single atom, three modes, and a maximum excitation number of two.}
    \label{fig:3modestates}
\end{figure}
In general, counting the set of possible states for multimode problems can be quite challenging, especially as the number of modes, the number of atoms, and the maximum excitation level increase. To tackle this issue, we have developed a combinatorial-based numerical program that generates all possible quantum states for a given number of few-level atoms (either two or three levels), a specified number of modes, and a defined maximum excitation number. For example, in Fig.~\ref{fig:3modestates}, we present a snapshot of the output of our program, showcasing the computed states for a single two-level atom, with three modes and two excitations. 

The zero-delay correlation function is derived by substituting Eq.~\eqref{Emulti} with \(N_m = 3\) into Eq.~\eqref{g2def}. By applying the dipole approximation — where we eliminate the \(x\)-dependence of the sine function and set its argument to a constant phase \(\phi_j \propto k_j\) — we obtain the following form of the correlation function for our trimode problem:
\begin{align}\label{g2multi}
    &g^{(2)}(0) =\nonumber\\
    &\frac{{\Bigg\langle} \left( \sum\limits^3_{j=1}\sqrt{\omega_j}\annihilation_j \sin{(\phi_j)}\right)^2\left(\sum\limits^3_{j=1}\sqrt{\omega_j}\creation_j \sin{(\phi_j)}\right)^2 \Bigg\rangle}{\left(\sum\limits^3_{j=1}[\sqrt{\omega_j}\annihilation_j \sin{(\phi_j)}]\sum\limits^3_{j=1}[\sqrt{\omega_j}\creation_j \sin{(\phi_j)]}\right)^2}.
\end{align}
As shown in Fig.~\ref{fig:3modestates}, the multimode problem with $N_m=3$ involves a total of fourteen basis states. This complexity makes the analytical form of the $g^{(2)}(0)$ function in this case quite challenging, so we will not present it here. Instead, in the next section, we will provide our numerical results for $g^{(2)}(0)$ and $g^{(2)}(\tau)$ for both single-mode and multi-mode cases.



\section{Results and Discussion} \label{sec:results}
In this section, we present our numerical results. We utilized the Quantum Toolbox in Python, commonly referred to as QuTiP \cite{qutip}, for our calculations. The numerical computations for the single mode case in QuTiP were carried out using the Hermitian Hamiltonian as given below, while setting $\hbar = 1$
\begin{equation}\label{singlemodeHHam}
    \hamiltonian = \Delta_{eg}\raising\lowering + \Delta_c\creation\annihilation + g(\raising\annihilation + \lowering\creation) + \varepsilon(\creation + \annihilation),
\end{equation}
and with the specification of the so-called collapse/jump operators as
\begin{align}\label{collapse1type1}
    \hat{C}_{a} =\sqrt{\kappa}\annihilation ~~~~\text{and}~~~~\hat{C}_{\sigma} =\sqrt{\gamma}\lowering,
\end{align}
in the master equation solver \texttt{qutip.steadystate} and the second-order correlation function calculator \texttt{qutip.coherence\_function\_g2}. On the other hand, for the multimode case, we extend the Hermitian Hamiltonian to take the form
\begin{align}
    &\hamiltonian = \Delta_{eg}\raising\lowering + \sum \limits^3_{i=1} \Big[\Delta_{ci}\creation_i\annihilation_i + g_i\left(\raising\annihilation_i + \lowering\creation_i\right)\nonumber \\ 
    &~~~~~~ + \varepsilon\left(\creation_i + \annihilation_i\right)\Big].
\end{align}
We set \( N_m = 3 \). In general, the two-photon collapse operators for any two modes \( i \) and \( j \) can be expressed as follows:
\begin{equation}\label{collapse1type22}
    \hat{C} = \hat{a}_i \hat{a}_j,~~~~\forall 1\leq i \leq N_m ~\text{and}~~\forall 1\leq j \leq N_m.
\end{equation}
To account for the two-photon dissipation process, we only consider collapse operators for cases where $i = j$. 

Before moving forward with our results, we give a quick refresher on the topic of quantum theory of photon statistics. For further details regarding non-classical states of light, we direct the interested reader to Ref.~\cite{fox2006quantum}. The second-order correlation function measures the degree of bunching between photons in a quantum optical system and characterizes the purity of a single-photon source. There are three distinct regimes of light described by the correlation function, each following unique statistics.
\begin{enumerate}
    \item \textit{Sub-Poissonian Light}: In this regime, \(g^{(2)} < 1\). Fock states and squeezed states are some important examples of this regime. Such light cannot be accurately described by classical electromagnetic theory \cite{gerry2023introductory}. It is important to note that the sub-Poissonian light and the antibunched light are not necessarily the same \cite{zou1990photon}. Light is considered antibunched when \(g^{(2)}(0) < g^{(2)}(\tau)\). 
    \vspace{-2mm}
    \item \textit{Poissonian Light}: When \(g^{(2)} = 1\), the light is Poissonian, representing an ideal scenario where light intensity is constant. The Glauber-Sudarshan coherent states are a key example of this regime \cite{gazeau2019coherent}. 
    \vspace{-2mm}
    \item \textit{Super-Poissonian Light}: In this regime, where \(g^{(2)} > 1\), the light is chaotic, such as thermal states \cite{fabre2020modes}, which display stochastic intensity fluctuations. 
\end{enumerate}
Single-photon sources belong to the sub-Poissonian regime and are generally antibunched. However, as mentioned earlier, some single-photon sources may exist in the bunched regime \cite{misiaszek2022applications}. The purer single-photon sources correspond to values of \(g^{(2)}\) smaller than 1.

\subsection{Single cavity mode}
As mentioned in Section \ref{sec:intro}, previous studies, such as Ref. \cite{zhang2023photon}, have explored the zero-delay second-order correlation function \( g^{(2)}(\tau = 0) \) for the single-mode cQED problem. These studies provide conclusive evidence that both the conventional and unconventional blockade optimal conditions result in sub-Poissonian light generated from a coherent input. However, it remains unclear whether this trend continues at non-zero delays. In this work, we first confirm the zero-delay findings previously reported and then present new numerical results for non-zero-delay scenarios. For our numerical simulations, unless otherwise specified, we use the parameters: \( \kappa = \gamma \) and \( g = 10\gamma \).


\subsubsection{Zero-delay problem}
\begin{figure*}
\centering
\begin{tabular}{cccc}
\includegraphics[width=3in, height=2.25in]{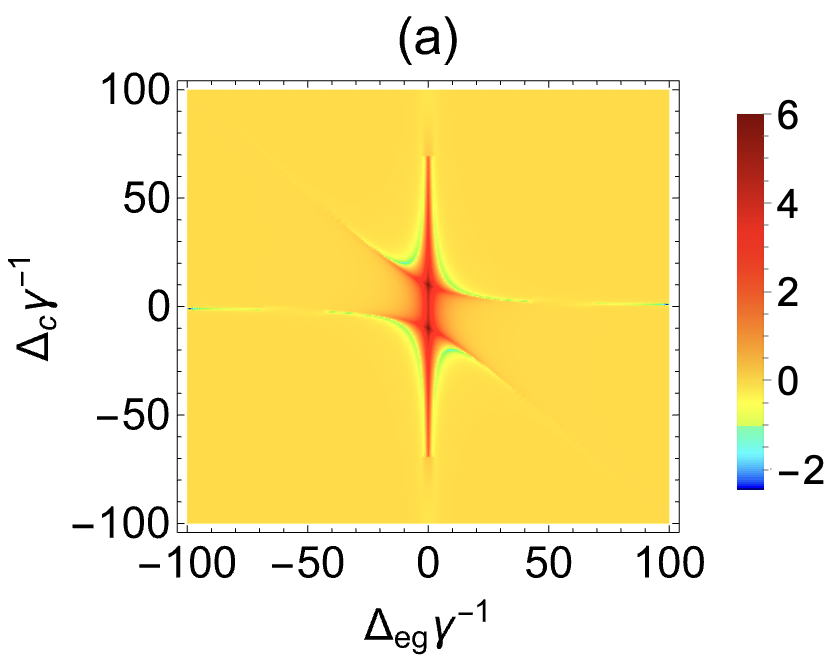} &
\hspace{5mm}\includegraphics[width=3in, height=2.25in]{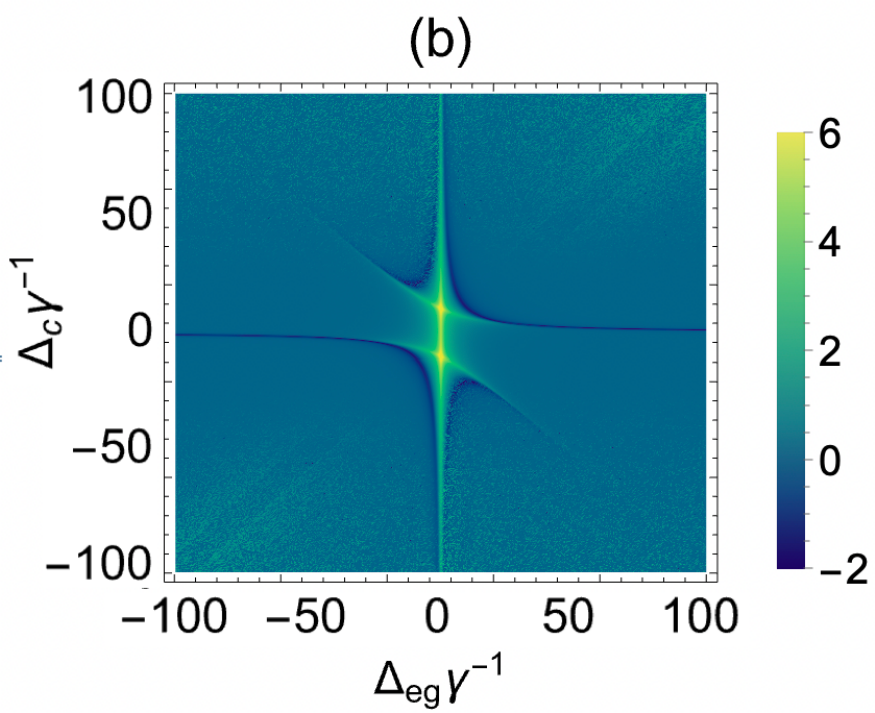} 
\end{tabular}
\captionsetup{
format=plain,
margin=1em,
justification=raggedright,
singlelinecheck=false
}
    \caption{(Color online) The density plot of $\log(g^{(2)}(0))$ is presented for the variable atom-drive detuning $\Delta_{eg}$ and the atom-cavity detuning $\Delta_c$. This is done in two ways: (a) using the analytic form derived in Eq.~\eqref{singlemodeg2analytic} and (b) utilizing a numerical  simulation based on collapse operators in QuTiP. To address potential accumulating numerical errors in our QuTiP code and to fulfill weak driving conditions, we have set $\varepsilon = 0.005\gamma$. Both detunings have been expressed in units of $\gamma$ such that $\Delta_{eg}\gamma^{-1}$ and $\Delta_{c}\gamma^{-1}$ are dimensionless.}
    \label{fig:singlemode}
\end{figure*}
We present our findings on the second-order correlation function \( g^{(2)}(0) \) for the single-mode case, as shown in Fig.~\ref{fig:singlemode}. Since \( g^{(2)}(0) \) can vary significantly, we have plotted \( \log(g^{(2)}(0)) \) against atomic detuning \( \Delta_{eg} \) and cavity detuning \( \Delta_c \) on a density plot, with both detunings measured in units of \( \gamma \). 

In Fig.~\ref{fig:singlemode}(a), we display the plot based on our analytical results from Eq.~\eqref{singlemodeg2analytic}, while Fig.~\ref{fig:singlemode}(b) illustrates the corresponding numerical results obtained from QuTiP using two-photon truncation. The QuTiP graphs were generated by consecutively applying the functions \texttt{qutip.steadystate()} and \texttt{qutip.coherence\_function\_g2()}. This process began with the initialization of the Hermitian Hamiltonian described in Eq.~\eqref{singlemodeHHam} and the associated list of collapse operators defined in Eqs.~\eqref{collapse1type1} and~\eqref{collapse1type22}. 

To significantly reduce computational errors arising from the small values generated by QuTiP during steady-state calculation, we used the Spirtes-Glymour-Scheines (or SGS) algorithm \cite{spirtes2000causation}. This algorithm was implemented to address potential erroneous negative eigenvalues of the density matrix produced by \texttt{steadystate()} by applying the method \texttt{qobj.trunc\_neg(method='sgs')}. During optimization of $\varepsilon$, we discovered that, under the SGS algorithm, setting $\varepsilon = 0.001\gamma$ significantly reduces the area of consistent and error-free results. As a result, we have chosen $\varepsilon = 0.001\gamma$ as the lower limit of the driving strength in the subsequent numerical calculations for this work.

Our initial observation reveals a notable similarity between Fig. \ref{fig:singlemode}(a) and Fig. \ref{fig:singlemode}(b). This similarity indicates a strong agreement between the analytical and numerical results under the weak drive assumption. Furthermore, both plots exhibit minimal values for the Conventional Photon Blockade (CPB) in the first and third quadrants, as well as for the (Unconventional Photon Blockade) UCPB in the second and fourth quadrants. For example, the dark blue hyperbolic regions in Fig.~\ref{fig:singlemode}(b) and yellowish-cyan regions in Fig.~\ref{fig:singlemode}(a) highlight these optimal conditions. 

\begin{figure}[]
    \centering
    \includegraphics[width=2.75in, height=1.75in]{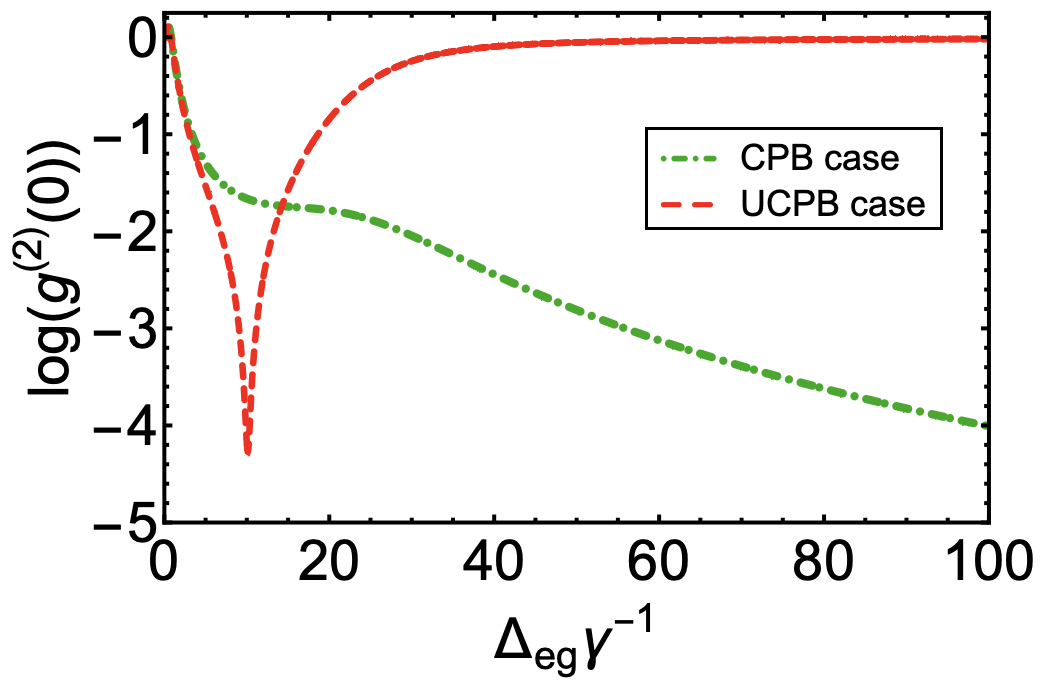} 
\captionsetup{
format=plain,
margin=1em,
justification=raggedright,
singlelinecheck=false
}
    \caption{(Color online) This plot presents $\log({g^{(2)}(0)})$ as a function of $\Delta_{eg}$. The green dotted-dashed curve corresponds to the CPB case, where the condition $\Delta_c = g^2 / \Delta_{eg}$ is satisfied. The red-dashed curve represents the UCPB scenario, which follows the condition $\Delta_c = -g^2 / \Delta_{eg} - \Delta_{eg}$.}
    \label{fig:optimalconditions}
\end{figure}

To further analyze our results, we plot \(\log g^{(2)}(0)\) in Fig.~\ref{fig:optimalconditions}, calculated using Eq.~\eqref{singlemodeg2analytic}, as a function of \(\Delta_{eg}\). We have selected optimal conditions for the CPB represented by the green dotted-dashed curve and for the UCPB represented by the red dashed curve. Our findings indicate that in the UCPB case \(\log g^{(2)}(0)\) reaches a sharp minimum at \(\Delta_{eg} \sim 10\gamma\). In the CPB scenario, the zero delay correlation function continues to decrease for larger atomic detunings (that is, when \(\Delta_{eg} \gg \gamma\)). In contrast, for the UCPB, \(g^{(2)}\) returns to the Poissonian regime as \(\Delta_{eg}\) increases.

\subsubsection{Nonzero-Delay Problem}
\begin{figure*}
\centering
\begin{tabular}{cccc}
\includegraphics[width=2.25in, height=1.5in]{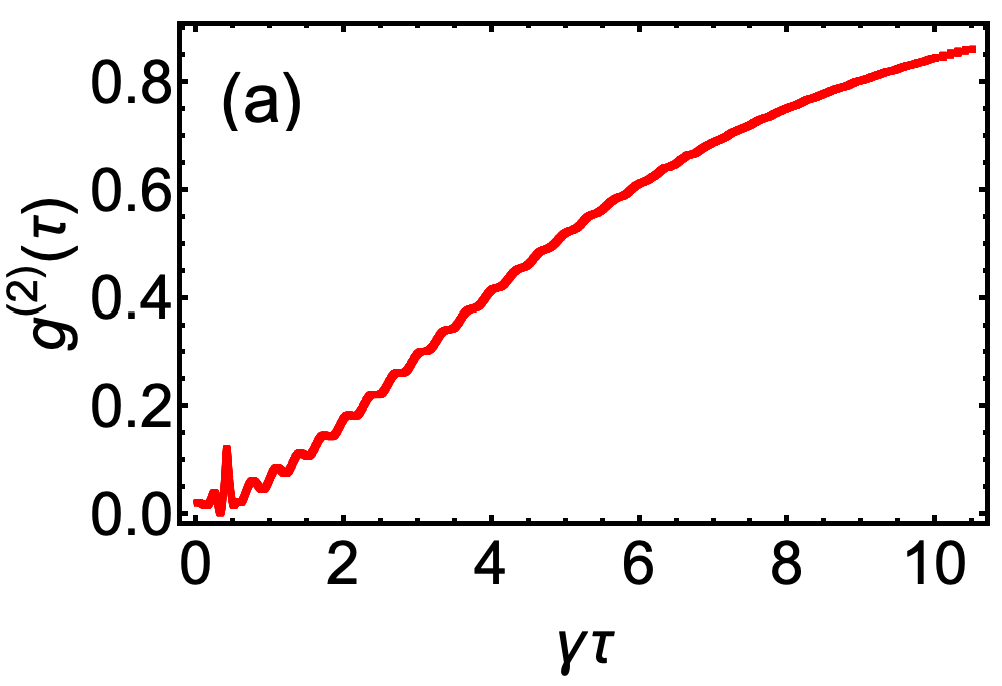} &
\hspace{-2mm}\includegraphics[width=2.35in, height=1.5in]{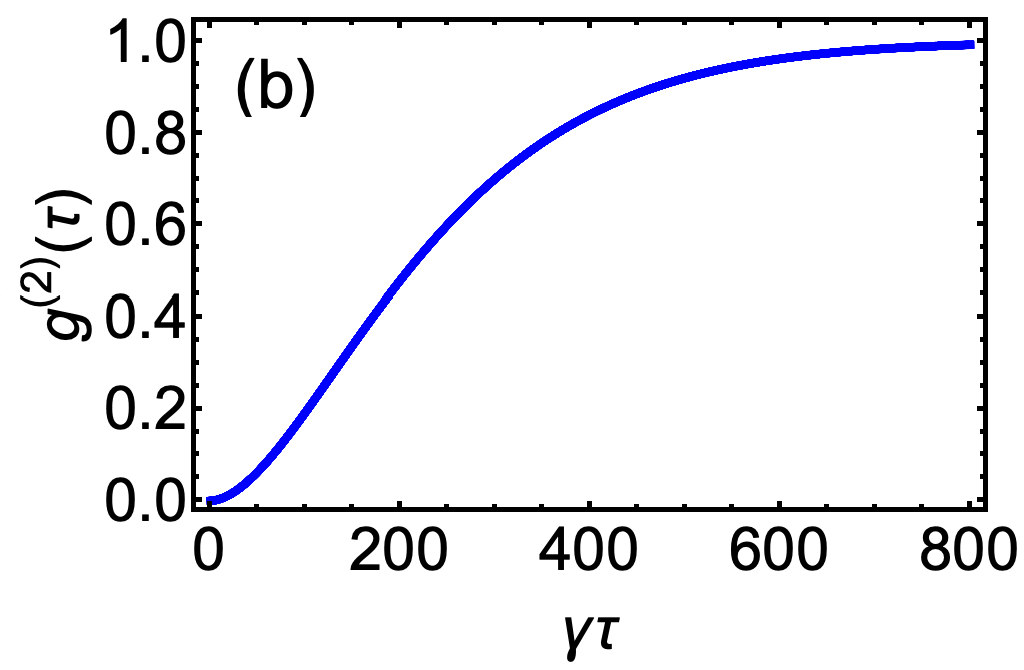} &
\hspace{-2mm}\includegraphics[width=2.35in, height=1.5in]{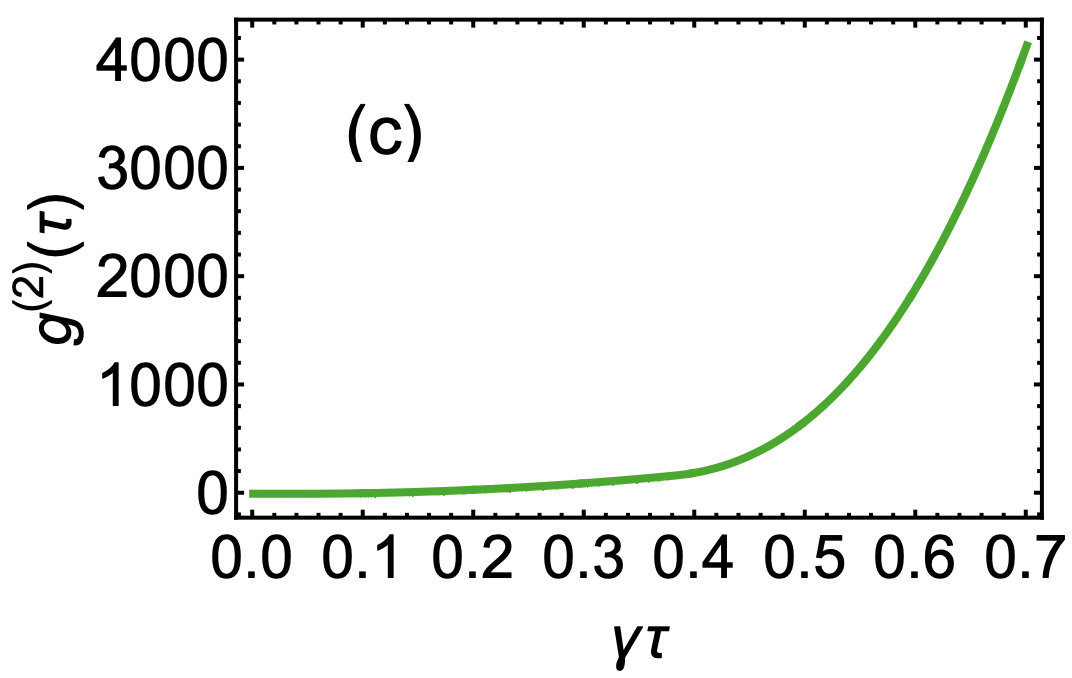} 
\end{tabular}
\captionsetup{
format=plain,
margin=1em,
justification=raggedright,
singlelinecheck=false
}
\caption{(Color online) This panel displays plots of our $g^{(2)}$ results as a function of the delay parameter $\tau$, measured in units of $\gamma^{-1}$. The conditions are as follows: (a) for the optimal condition of the CPB with parameters $(\Delta_{eg}, \Delta_c) = (10, 10)\gamma$ and $\varepsilon = 0.00172\gamma$, (b) for the CPB optimal condition with parameters $(\Delta_{eg}, \Delta_c) = (100, 1)\gamma$ with $\varepsilon = 0.00172$, and (c) for the near-UCPB optimal condition with $(\Delta_{eg}, \Delta_c) = (8.9618, -19.4236)\gamma$ and $\varepsilon = 0.005\gamma$.}
\label{fig:taucpb1}
\end{figure*}
For the case of nonzero delay, even in the single-mode scenario, obtaining an analytic solution is quite challenging. Therefore, we once again relied on a numerical simulation executed in QuTiP where we calculated \( g^{(2)}(\tau) \) (as defined in Eq.~\eqref{g2def}) over a wide range of \( \tau \) values. As before, we selected \( \varepsilon = 10^{-3}\gamma \), which is small enough to remain within the weak driving regime but large enough to ensure that computational errors in \( g^{(2)}(\tau) \) would be negligible.

\begin{table}[h!]
    \centering
    \begin{tabular}{c|c|c|c}
       $(\Delta_{eg},\Delta_c)/\gamma$ & $\varepsilon / \gamma$ & Analytic $ g^{(2)}(0)$  & Numeric $g^{(2)}(0)$ \\
      \hline (10,10) & 0.001 & 0.0220 & 0.0220 \\
       & 0.005 &  & 0.0220 \\  
       & 0.01 &  & 0.0221 \\
      (100,1) & 0.001 & $9.9978\times 10^{-5}$ & $1.0806\times 10^{-4}$\\
       & 0.005 &  & $3.0191\times 10^{-4}$\\
        & 0.01 &  & $9.0728\times 10^{-4}$\\
      (10,-20) & 0.001 & $5.5068\times 10^{-5}$ & $5.5108\times 10^{-5}$ \\
       & 0.005 &  & $0.012930$ \\
       & 0.01 &  & $0.0029752$ \\
      (50,-52) & 0.001 & 0.9042 & 1831.84 \\
       & 0.005 &  & 2.1262 \\
       & 0.01 &  & 0.9502 \\
      (50,50) & 0.001 & 0.9983 & 624.15 \\ 
       & 0.005 &  & 1.0003 \\ 
       & 0.01 &  & 0.9983 \\
       (50,0) & 0.001 & 1.0132 & 1.0133 \\
        & 0.005 &  & 1.0132 \\
       & 0.01 &  & 1.0131 
    \end{tabular}
    \captionsetup{
format=plain,
margin=1em,
justification=raggedright,
singlelinecheck=false
}
    \caption{The function $g^{(2)}(0)$ is analyzed for selected values of ($\Delta_{eg}, \Delta_c$) with driving strengths in the range of $0.001\gamma \leq \varepsilon \leq 0.01\gamma$. The approximation used to derive Eq.~\eqref{singlemodeg2analytic} is in good agreement with numerical calculations, except for cases involving large cavity detunings at the lowest driving strengths.}
    \label{tab:selected}
\end{table}

To validate our results, we compared our non-zero delay findings by setting \(\tau = 0\) for each driving strength at the specified \((\Delta_{eg}, \Delta_c)\) values to the analytical result, which is independent of driving strength, for the same pair of detunings. The values of the correlation function for selected parameters are presented in Table~\ref{tab:selected}. From Table~\ref{tab:selected}, we indicate that the optimal driving strength for achieving the best agreement between analytical and numerical results lies between $0.001\gamma$ and $0.005\gamma$. In particular, the numerical results exhibit the greatest margin for error at smaller driving strengths when there are high cavity detunings. In contrast, increasing the driving strength to compensate for this error results in smaller values of $g^{(2)}(0)$ (for example, at $(\Delta_{eg}, \Delta_c) = (10, -20)\gamma$), leading to less antibunching as the weak driving approximation begins to break down. Therefore, it is crucial to identify intermediate $\varepsilon$ values that minimize computational error while still maintaining the validity of the weak driving assumption.

In Fig.~\ref{fig:taucpb1}(a), we plot \( g^{(2)}(\tau) \) for \( (\Delta_a, \Delta_c) = (10, 10)\gamma \) and \( \varepsilon = 0.00172\gamma \) under conditions that meet the optimal criteria for the CPB. We observe that the correlation function starts deep within the sub-Poissonian regime and increases as \( \tau \) increases. Although there are slight oscillations at first, as \( \tau \) becomes larger, the function shows a clear trend toward asymptotically approaching \( g^{(2)}(\tau) = 1 \). Intuitively, this behavior suggests that the probability of detecting a second photon, given that we have already detected one, increases with the delay between the two phodetection events. Furthermore, we have strong evidence of antibunching demonstrated by the condition that \( g^{(2)}(0) < g^{(2)}(\tau) \). As we approach very large values of \( \tau \) (for example, when \( \tau \geq 10\gamma^{-1} \)), we see that \( g^{(2)}(\tau) \) approaches 1, indicating a behavior similar to that of a Poissonian light or a coherent state. 

In Fig.~\ref{fig:taucpb1}(b), we maintain the same parameters and conditions as in Fig.~\ref{fig:taucpb1}(a), with the exception of the optimal conditions for the CPB. Specifically, we adjust the detunings to new values of \((\Delta_{eg}, \Delta_c) = (100, 1)\gamma\). We observe an overall similarity in behavior compared to Fig.~\ref{fig:taucpb1}(a). However, due to the larger parameter \(\Delta_{eg}\) in this context, we find that the growth of \(g^{(2)}(\tau)\) is relatively slow before it reaches the Poissonian unit value.  

\begin{figure*}
\centering
\begin{tabular}{cccc}
\includegraphics[width=3in, height=2.25in]{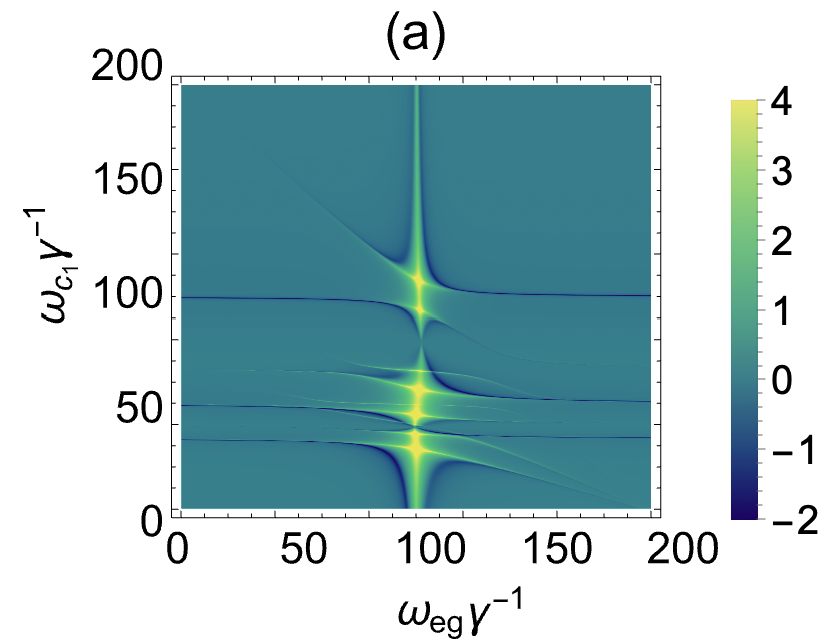} &
\hspace{2mm}\includegraphics[width=3in, height=2.25in]{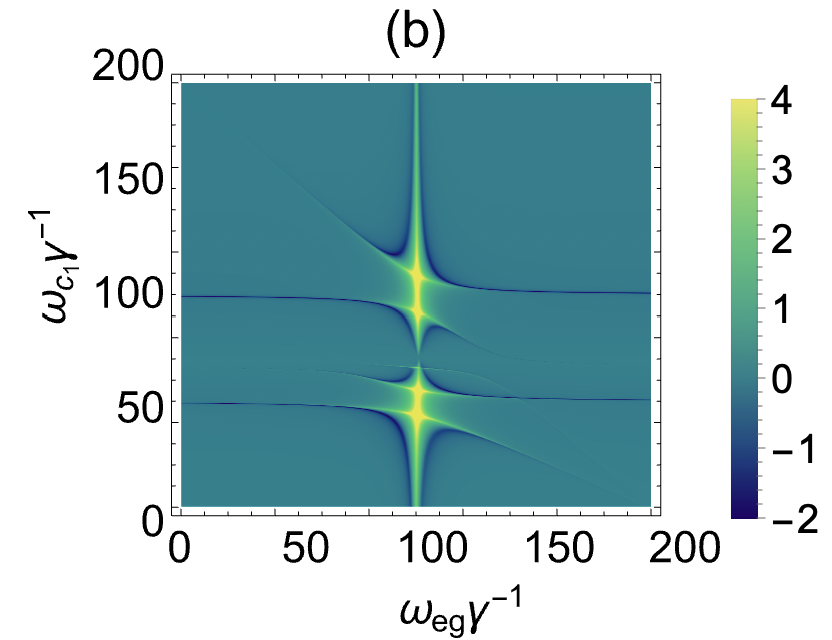} 
\end{tabular}
\captionsetup{
format=plain,
margin=1em,
justification=raggedright,
singlelinecheck=false
}
    \caption{(Color online) The plot of $\log({g^{(2)}})$ under zero-delay conditions is shown for three modes with variable $\omega_{eg}$ and $\omega_1$, using $\varepsilon = 0.001\gamma$. The key difference between plots (a) and (b) is that in plot (b), the value of $k_1 x$ has been adjusted to $\frac{\pi}{3}$, allowing for the decoupling of the third cavity harmonic from the atom.}
    \label{fig:threemode_varwcwa}
\end{figure*}

Finally, Fig.~\ref{fig:taucpb1}(c), illustrates \( g^{(2)}(\tau) \) for a point that is approximately located on the UCPB optimal condition curve. This particular point was chosen to have the smallest cavity detuning in order to minimize numerical errors and to be close to the atomic detuning associated with the minimum shown in Fig.~\ref{fig:optimalconditions}. The driving strength parameter \( \varepsilon = 0.005 \) satisfies the criteria for the weak driving approximation, as demonstrated by a 0.4\% difference between the analytic and numeric values of \( g^{(2)}(0) \) for this set of parameters (with \( g^{(2)}(0) = 0.03866 \) and \( g^{(2)}(0) = 0.03907 \), respectively). It is evident that the correlation function in this case increases monotonically at a much faster rate compared to the optimal CPB condition. This observation suggests that utilizing UCPB may produce single photons with significantly less temporal spacing than using CPB.

\subsection{Multimode cavity modes}
We will now focus on the multimode cavity problem. As a practical example, we consider an optical cavity that supports three optical modes. We will follow the theoretical framework described in Section II-B. For our parameters, we use Eq.~\(\eqref{Eq10purane}\) to establish the interaction strengths, setting \(g_1 = g_0 \sin(k_1 x)\), where \(k_1 x\) is considered a constant under the dipole approximation, and \(g_0\) represents the base interaction strength. Unless otherwise specified, we assume \(\kappa = \gamma = 1\) (with \(\kappa\) being the same for each mode), set \(g_0 = 10\gamma\), \(k_1 x = \frac{\pi}{4}\), and \(\omega_d = 100\gamma\). 

In this multimode scenario, we can no longer consider atomic and cavity detunings as independent variables. Instead, we will adjust two parameters based on the driving frequency, the fundamental cavity mode frequency, and the atomic resonant frequency, while keeping the third parameter constant. For this discussion, we have chosen \(\omega_d\) as a constant frequency. We will now discuss our multimode results, following the same structure as in the single-mode case. First, we will examine the case of zero delay, followed by the case of non-zero delay.

\subsubsection{Zero-Delay Problem}
\begin{figure*}
\centering
\begin{tabular}{cccc}
\includegraphics[width=2.25in, height=1.5in]{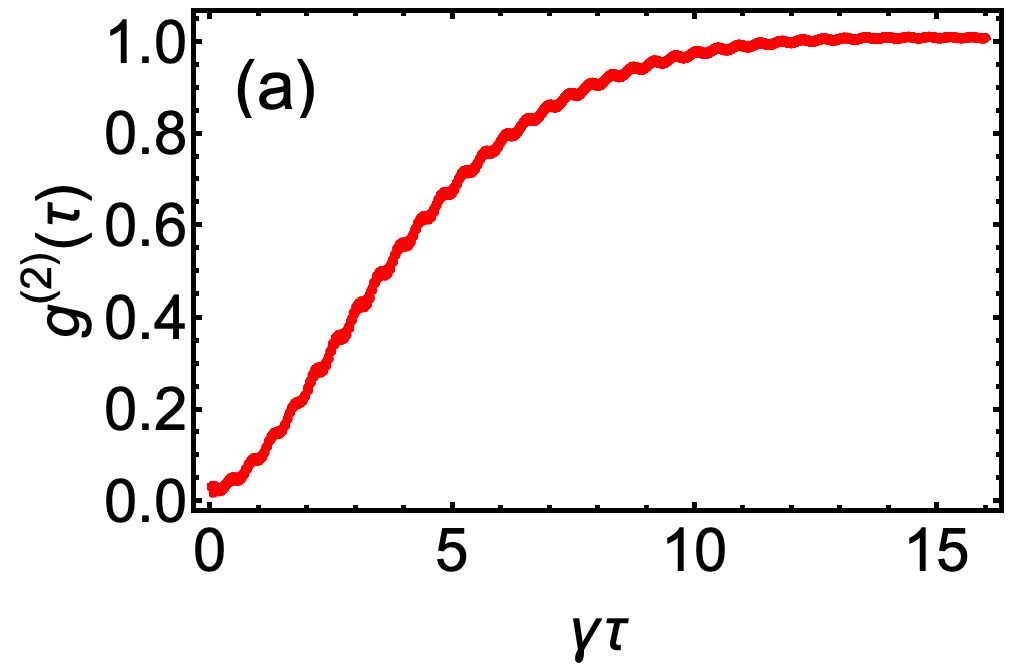} &
\hspace{-2mm}\includegraphics[width=2.35in, height=1.5in]{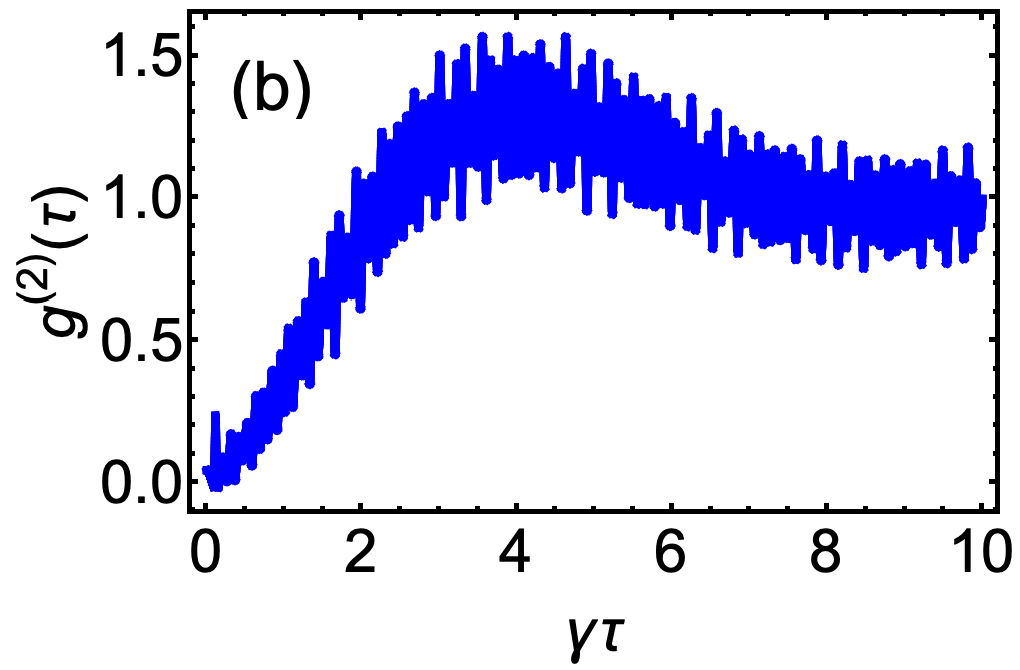} &
\hspace{-2mm}\includegraphics[width=2.3in, height=1.54in]{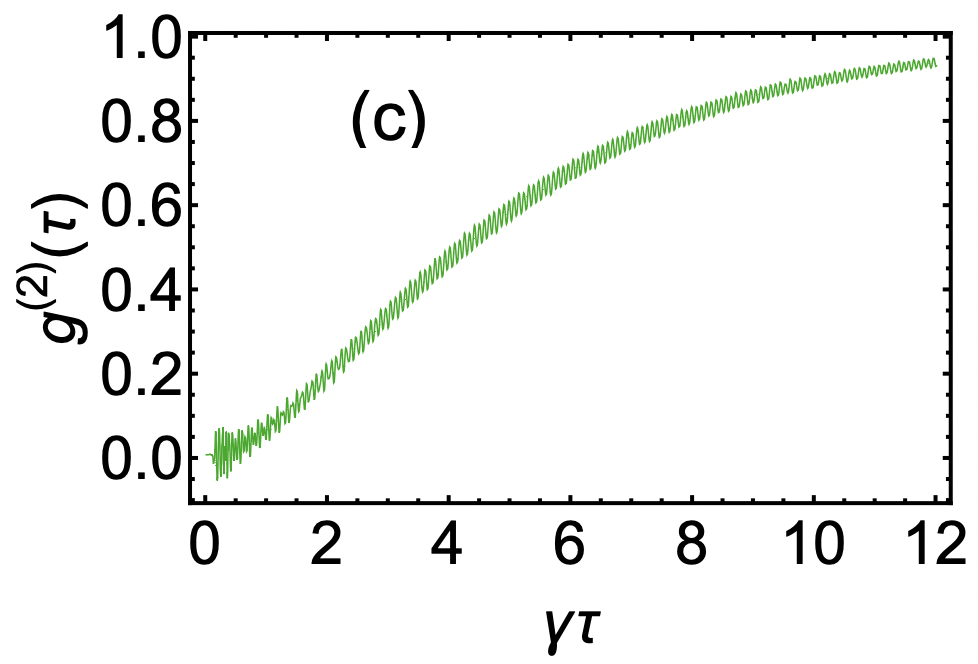} 
\end{tabular}
\captionsetup{
format=plain,
margin=1em,
justification=raggedright,
singlelinecheck=false
}
    \caption{(Color online) The above plots illustrate the variation of \( g^{(2)} \) as a function of \( \tau \) for the multimode case: (a) at CPB optimal conditions with parameters \( (\omega_{eg}, \omega_{c_1}) = (108.5, 108.5)\gamma \), (b) at near-UCPB conditions for the second harmonic with parameters \( (\omega_{eg}, \omega_{c_1}) = (90, 23)\gamma \), and (c) at near-CPB conditions for the third harmonic with parameters \( (\omega_{eg}, \omega_{c_1}) = (112.5, 58.5)\gamma \). All other parameters remain consistent with previous analyses. }
    \label{fig:taucpb}
\end{figure*}
In the case of three modes, we calculate the zero-delay correlation function using Eq.~\(\eqref{g2multi}\) within the limit of up to two excitations. The resulting behavior is illustrated as a density plot based on the frequencies \(\omega_{eg}\) and \(\omega_{c_1}\). To obtain the steady-state probability amplitudes required for this calculation, we substituted the general three-mode quantum state into the time-dependent Schrödinger equation, which led to a system of fourteen coupled differential equations which we numerically solved. Figure~\ref{fig:3modestates} displays the set of basis states corresponding to these amplitudes.

Figure \ref{fig:threemode_varwcwa}(a) shows the function \( g^{(2)}(0) \) for three modes with fixed \(\omega_d\) and \(\varepsilon = 0.001\gamma\). The overall shape of this graph resembles that of Figure \ref{fig:singlemode}; however, it now contains three distinct regions exhibiting super-Poissonian statistics, along with three pairs of hyperbolic sub-Poissonian regions. Notably, there is significant overlap between these two types of regions at a lower fundamental frequency of the cavity. The first sub-Poissonian region is symmetric around \(\omega_{c_1} = 100\gamma\), the second around \(\omega_{c_1} = 50\gamma\), and the third is symmetric about \(\omega_{c_1} \approx 33\gamma\). These frequencies correspond to the values at which the first, second, and third cavity modes resonate with the driving field. This resonance explains why the lower half of the density plot exhibits overlap, while the upper region is more distinct. 

As the harmonic number of the cavity that resonates with the driving field increases in integer steps, the corresponding fundamental frequency decreases by a smaller amount. For example, the fourth harmonic of a multimode cavity resonates with the drive at \(\omega_{c_1} = 25\gamma\), the fifth at \(\omega_{c_1} = 20\gamma\), the sixth at \(\omega_{c_1} = 16.667\gamma\), and so on. This trend suggests that examining more cavity modes beyond three may lead to significantly lower values of \(g^{(2)}(0)\), where multiple regions of sub-Poissonian statistics overlap. Additionally, the presence of several small \(g^{(2)}\) regions with larger widths implies that the physical implementations of single-photon sources within this system will be more flexible. This versatility allows for the detection of single photons across a wider range of parameter combinations.


Figure~\ref{fig:threemode_varwcwa}(b) supports the argument that the resonance of the first, second, and third cavity harmonics with the driving field accounts for the three single-mode-like regions observed in the \( g^{(2)}(0) \) graph for the three modes. By setting \( k_1 x = \frac{\pi}{3} \); a process that can be achieved experimentally by adjusting the position of the atom within the cavity; we effectively decouple the atom from the third cavity mode. As a result, we see the elimination of the single-mode-like region for \( \omega_1 = 33 \). This leads us to conclude that each mode in the model contributes its own region of sub- or super-Poissonian statistics to the zero-delay correlation function.

\subsubsection{Nonzero-Delay Problem}
To calculate the non-zero delay correlation function and examine the presence or absence of antibunching in the three-mode case, we extended the numerical QuTiP program used for the single-mode case. Instead of directly calculating the correlation function with the annihilation operator, we utilized the electric field operator for three modes. We defined our collapse operators according to Eq.~\eqref{collapse1type22}, where \(i = j\).

\begin{table}[]
    \centering
    \begin{tabular}{c|c|c|c}
       ($\omega_{eg}$, $\omega_{c_1}$)/$\gamma$  & $\varepsilon/\gamma$ & Analytic $g^{(2)}(0)$ & Numerical $g^{(2)}(0)$ \\
      \hline  (108.5,108.5) & 0.001 & 0.0338 & 0.187 \\
      & 0.005 & & 0.0326 \\
      & 0.01 & & 0.0326 \\
      (90, 23) & 0.001 & 0.0389 & 0.0411 \\
      & 0.005 & & 0.0414 \\
      & 0.01 & & 0.0411 \\
      (112.5, 58.5) & 0.001 & 0.0120 & 0.0162 \\
      & 0.005 & & 0.0100 \\
      & 0.01 & & 0.0100 \\
      (50,100) & 0.001 & 0.6555 & 0.7971 \\
      & 0.005 & & 0.7971 \\
      & 0.01 & & 0.7971
    \end{tabular}
    \captionsetup{
format=plain,
margin=1em,
justification=raggedright,
singlelinecheck=false
}
    \caption{The values of $g^{(2)}(0)$ are presented for selected combinations of ($\omega_{eg}, \omega_{c_1}$) with driving strengths in the range of $0.001 \leq \varepsilon \leq 0.01$. The approximations used for the steady-state analysis, obtained by solving the coupled differential equations for amplitudes, show good agreement with the master equation-based numerical calculations conducted using QuTiP.}
    \label{tab:threemodeselected}
\end{table}

We begin by examining whether the function \(g^{(2)}\) produces consistent values between the method used to solve the coupled differential equations for the amplitudes and the numerical calculations based on the master equation, both performed at the initial time point \(\tau=0\). To facilitate this comparison, we have included Table~\ref{tab:threemodeselected}, which presents the values of \(g^{(2)}(0)\) for specific combinations of \(\omega_{eg}\), \(\omega_{c_1}\), and \(\varepsilon\) according to the two methods. For this analysis, we set \(k_1x = \frac{\pi}{4}\).

The pair \((\omega_{eg}, \omega_{c_1})/\gamma = (108.5, 108.5)\) corresponds to the near-CPB condition for the fundamental mode. The pair \((90, 23)\) relates to the near-optimal UCPB condition for the third harmonic, while \((112.5, 58.5)\) reflects the near-optimal CPB condition for the second harmonic. Lastly, the pair \((50, 100)\) represents an arbitrarily selected point situated outside the sub/super-Poissonian structure but still within an acceptable range of computational error. The values presented in Table~\ref{tab:threemodeselected} demonstrate general consistency in the correlation function in the three driving strengths listed. In particular, the range of \(0.001 \leq \varepsilon \leq 0.005\) shows the highest degree of agreement. Consequently, we have chosen \(\varepsilon = 0.0025\) to generate the subsequent plots of \(g^{(2)}(\tau)\).

Figure~\ref{fig:taucpb}(a) presents the graph of \( g^{(2)}(\tau) \) for a specific point under near-CPB optimal conditions, which is slightly beyond the strongest minimal region for the fundamental mode. Similarly to the single-mode scenario, we observe small oscillations superimposed on a general increasing trend in the graph. This trend ultimately reaches a value of one over an extended time scale (approximately \( \tau \geq 10\gamma^{-1} \)). However, in the multimode case, we note a faster rate at which \( g^{(2)}(\tau) \) approaches this unit value compared to the single-mode problem. Overall, this behavior demonstrates that the single photons emitted from the multimode cQED setup are indeed antibunched and exhibit sub-Poissonian statistics.

Figure~\ref{fig:taucpb}(b) shows the results for \( g^{(2)}(\tau) \) at a frequency pair close to optimal for UCPB in the second harmonic. The second-order correlation function reaches a Poissonian regime faster than in the CPB case, but not as quickly as in the single-mode scenario (see Fig.~\ref{fig:taucpb1}(c) for comparison). This result indicates that higher modes strengthen the photon blockade effect and support the presence of antibunching. Rapid oscillations in the data, approximately 0.2, are likely due to the chosen step size $\Delta\tau$ for the horizontal scale ($\Delta\tau = 0.01\gamma^{-1}$). It was not possible to increase the number of time steps because of computational limitations, so this aspect needs further study.

Finally, Fig.~\ref{fig:taucpb}(c) illustrates the nonzero delay correlation function for a frequency pair that is roughly aligned with the CPB's optimal conditions for the third harmonic. As seen, achieving Poissonian statistics takes longer than in Fig.~\ref{fig:taucpb}(a). The function exhibits rapid oscillations, eventually approaching \( g^{(2)}(\tau) = 1 \). This suggests that the photon blockade caused by resonance with one harmonic in a multimode scenario behaves similarly to that in a single-mode scenario. Once again, we have demonstrated antibunching under these optimal conditions. 


\section{Summary and Conclusions}\label{sec:conclusion}
In this paper, we have explored both conventional and unconventional photon blockade in multimode cQED systems, with a focus on generating single photons for possible quantum information science applications. The study employs a weak driving limit, which allows the Hilbert space to be restricted to two excitations for numerical analysis, and considers the strong coupling regime of cQED.

For the single-mode problem, the analytic and numerical results demonstrated strong agreement, thus confirming previously reported findings \cite{zhang2023photon}. The minimal values $g^{(2)}(0)$ formed hyperbolic patterns for the CPB and the UCPB in the odd and even quadrants, respectively, of the density plot of $\log(g^2(0))$ versus $\Delta_{eg}$ and $\Delta_{c}$. As a novel extension, $g^{(2)}$ was numerically evaluated as a function of $\tau$. Under optimal conditions of CPB with $(\Delta_{eg},\Delta_c) = (10, 10)\gamma$, the system initially exhibited deep Poissonian statistics, with $g^{(2)}(\tau)$ increasing with $\tau$ and approaching unity for $\tau > 10\gamma^{-1}$. In contrast, the parameter set $(\Delta_{eg},\Delta_c) = (100, 1)\gamma$ produced a similar profile, but with a slower increase in $g^{(2)}(\tau)$.

The multimode problem is examined using the same parameter considerations as previously, except with $k_1x=\pi/4$ and $\omega_d=100\gamma$. In the tri-modal scenario, three distinct regions exhibiting super-Poissonian and sub-Poissonian behavior are observed. The sub-Poissonian regions are symmetric about three cavity mode frequencies that are resonant with the driving field. This trend suggests that increasing the number of modes generates multiple regions of sub-Poissonian statistics, thereby providing greater flexibility in the application of multimode cQED setups compared to the single-mode case. In the non-zero delay scenario, under near-CPB optimal conditions, the $g^{(2)}(\tau)$ function increases at a faster rate than in the single-mode case. When the same CPB optimal conditions are applied to the third harmonic, the curve's behavior resembles that of the single-mode case, but with a slower rate of convergence to $g^{(2)}(\tau)=1$. Future work will extend the number of modes considered and move beyond the weak-driving assumption to determine whether single-photon generation can be more effectively engineered.


\section*{Acknowledgements} \label{sec:acknowledgements}
We are thankful to Wenchao Ge for helpful comments on the manuscript. IMM acknowledges financial support from the NSF LEAPS-MPS Grant \# 2212860. CM would like to gratefully acknowledge support from the Miami University Summer Scholars program.


\bibliographystyle{unsrt}
\bibliography{paper}

\begin{thebibliography}{10}

\bibitem{lounis2005single}
Brahim Lounis and Michel Orrit.
\newblock Single-photon sources.
\newblock {\em Reports on Progress in Physics}, 68(5):1129, 2005.

\bibitem{o2009photonic}
Jeremy~L O'brien, Akira Furusawa, and Jelena Vu{\v{c}}kovi{\'c}.
\newblock Photonic quantum technologies.
\newblock {\em Nature Photonics}, 3(12):687--695, 2009.

\bibitem{couteau2023applications}
Christophe Couteau, Stefanie Barz, Thomas Durt, Thomas Gerrits, Jan Huwer,
  Robert Prevedel, John Rarity, Andrew Shields, and Gregor Weihs.
\newblock Applications of single photons to quantum communication and
  computing.
\newblock {\em Nature Reviews Physics}, 5(6):326--338, 2023.

\bibitem{knill2001scheme}
Emanuel Knill, Raymond Laflamme, and Gerald~J Milburn.
\newblock A scheme for efficient quantum computation with linear optics.
\newblock {\em Nature}, 409(6816):46--52, 2001.

\bibitem{beveratos2002single}
Alexios Beveratos, Rosa Brouri, Thierry Gacoin, Andr{\'e} Villing,
  Jean-Philippe Poizat, and Philippe Grangier.
\newblock Single photon quantum cryptography.
\newblock {\em Physical Review Letters}, 89(18):187901, 2002.

\bibitem{mirza2013single}
Imran~M Mirza, SJ~van Enk, and HJ~Kimble.
\newblock Single-photon time-dependent spectra in coupled cavity arrays.
\newblock {\em Journal of the Optical Society of America B}, 30(10):2640--2649,
  2013.

\bibitem{couteau2018spontaneous}
Christophe Couteau.
\newblock Spontaneous parametric down-conversion.
\newblock {\em Contemporary Physics}, 59(3):291--304, 2018.

\bibitem{zhang2021spontaneous}
Chao Zhang, Yun-Feng Huang, Bi-Heng Liu, Chuan-Feng Li, and Guang-Can Guo.
\newblock Spontaneous parametric down-conversion sources for multiphoton
  experiments.
\newblock {\em Advanced Quantum Technologies}, 4(5):2000132, 2021.

\bibitem{karan2020phase}
Suman Karan, Shaurya Aarav, Homanga Bharadhwaj, Lavanya Taneja, Arinjoy De,
  Girish Kulkarni, Nilakantha Meher, and Anand~K Jha.
\newblock Phase matching in $\beta$-barium borate crystals for spontaneous
  parametric down-conversion.
\newblock {\em Journal of Optics}, 22(8):083501, 2020.

\bibitem{bock2016highly}
Matthias Bock, Andreas Lenhard, Christopher Chunnilall, and Christoph Becher.
\newblock Highly efficient heralded single-photon source for telecom
  wavelengths based on a ppln waveguide.
\newblock {\em Optics Express}, 24(21):23992--24001, 2016.

\bibitem{oxborrow2005single}
Mark Oxborrow and Alastair~G Sinclair.
\newblock Single-photon sources.
\newblock {\em Contemporary Physics}, 46(3):173--206, 2005.

\bibitem{scheel2009single}
Stefan Scheel.
\newblock Single-photon sources--an introduction.
\newblock {\em Journal of Modern Optics}, 56(2-3):141--160, 2009.

\bibitem{paul1982photon}
H~Paul.
\newblock Photon antibunching.
\newblock {\em Reviews of Modern Physics}, 54(4):1061, 1982.

\bibitem{lopez2022loss}
Juan~Camilo L{\'o}pez~Carre{\~n}o, Eduardo Zubizarreta~Casalengua, Blanca
  Silva, Elena del Valle, and Fabrice~P Laussy.
\newblock Loss of antibunching.
\newblock {\em Physical Review A}, 105(2):023724, 2022.

\bibitem{bassani2005encyclopedia}
Franco Bassani, Gerald~L Liedl, and Peter Wyder.
\newblock Encyclopedia of condensed matter physics.
\newblock 2005.

\bibitem{grabert2013single}
Hermann Grabert and Michel~H Devoret.
\newblock {\em Single charge tunneling: Coulomb blockade phenomena in
  nanostructures}, volume 294.
\newblock Springer Science \& Business Media, 2013.

\bibitem{birnbaum2005photon}
Kevin~M Birnbaum, Andreea Boca, Russell Miller, Allen~D Boozer, Tracy~E
  Northup, and H~Jeff Kimble.
\newblock Photon blockade in an optical cavity with one trapped atom.
\newblock {\em Nature}, 436(7047):87--90, 2005.

\bibitem{imamoḡlu1997strongly}
A~Imamoḡlu, Helmut Schmidt, Gareth Woods, and Moshe Deutsch.
\newblock Strongly interacting photons in a nonlinear cavity.
\newblock {\em Physical Review Letters}, 79(8):1467, 1997.

\bibitem{zhou2025universal}
Yan-Hui Zhou, Tong Liu, Qi-Ping Su, Xing-Yuan Zhang, Qi-Cheng Wu, Dong-Xu Chen,
  Zhi-Cheng Shi, HZ~Shen, and Chui-Ping Yang.
\newblock Universal photon blockade.
\newblock {\em Physical Review Letters}, 134(18):183601, 2025.

\bibitem{dayan2008photon}
Barak Dayan, AS~Parkins, Takao Aoki, EP~Ostby, KJ~Vahala, and HJ~Kimble.
\newblock A photon turnstile dynamically regulated by one atom.
\newblock {\em Science}, 319(5866):1062--1065, 2008.

\bibitem{hamsen2017two}
Christoph Hamsen, Karl~Nicolas Tolazzi, Tatjana Wilk, and Gerhard Rempe.
\newblock Two-photon blockade in an atom-driven cavity qed system.
\newblock {\em Physical Review Letters}, 118(13):133604, 2017.

\bibitem{flayac2017unconventional}
H~Flayac and V~Savona.
\newblock Unconventional photon blockade.
\newblock {\em Physical Review A}, 96(5):053810, 2017.

\bibitem{snijders2018observation}
HJ~Snijders, JA~Frey, J~Norman, H~Flayac, V~Savona, AC~Gossard, JE~Bowers,
  MP~Van~Exter, D~Bouwmeester, and W~L{\"o}ffler.
\newblock Observation of the unconventional photon blockade.
\newblock {\em Physical Review Letters}, 121(4):043601, 2018.

\bibitem{jaynes2005comparison}
Edwin~T Jaynes and Frederick~W Cummings.
\newblock Comparison of quantum and semiclassical radiation theories with
  application to the beam maser.
\newblock {\em Proceedings of the IEEE}, 51(1):89--109, 2005.

\bibitem{larson2021jaynes}
Jonas Larson and Themistoklis Mavrogordatos.
\newblock {\em The {J}aynes--{C}ummings model and its descendants: modern
  research directions}.
\newblock IoP Publishing, 2021.

\bibitem{zhang2023photon}
Haoliang Zhang and Zhenglu Duan.
\newblock Photon blockade in the {J}aynes-{C}ummings model with two-photon
  dissipation.
\newblock {\em Optics Express}, 31(14):22580--22593, 2023.

\bibitem{kristensen2014modes}
Philip~Tr{\o}st Kristensen and Stephen Hughes.
\newblock Modes and mode volumes of leaky optical cavities and plasmonic
  nanoresonators.
\newblock {\em ACS Photonics}, 1(1):2--10, 2014.

\bibitem{naik2017random}
RK~Naik, N~Leung, S~Chakram, Peter Groszkowski, Y~Lu, Nathan Earnest, DC~McKay,
  Jens Koch, and David~I Schuster.
\newblock Random access quantum information processors using multimode circuit
  quantum electrodynamics.
\newblock {\em Nature Communications}, 8(1):1--7, 2017.

\bibitem{filipp2011multimode}
Stefan Filipp, M~G{\"o}ppl, JM~Fink, M~Baur, R~Bianchetti, L~Steffen, and
  Andreas Wallraff.
\newblock Multimode mediated qubit-qubit coupling and dark-state symmetries in
  circuit quantum electrodynamics.
\newblock {\em Physical Review A}, 83(6):063827, 2011.

\bibitem{von2024engineering}
Uwe von L{\"u}pke, Ines~C Rodrigues, Yu~Yang, Matteo Fadel, and Yiwen Chu.
\newblock Engineering multimode interactions in circuit quantum
  acoustodynamics.
\newblock {\em Nature Physics}, 20(4):564--570, 2024.

\bibitem{rumi2000structure}
Mariacristina Rumi, Jeffrey~E Ehrlich, Ahmed~A Heikal, Joseph~W Perry, Stephen
  Barlow, Zhongying Hu, Dianne McCord-Maughon, Timothy~C Parker, Harald
  R{\"o}ckel, Sankaran Thayumanavan, et~al.
\newblock Structure- property relationships for two-photon absorbing
  chromophores: bis-donor diphenylpolyene and bis (styryl) benzene derivatives.
\newblock {\em Journal of the American Chemical Society}, 122(39):9500--9510,
  2000.

\bibitem{van2002optical}
Vien Van, Tarek~A Ibrahim, Philippe~P Absil, Fred~G Johnson, Rohit Grover, and
  P-T Ho.
\newblock Optical signal processing using nonlinear semiconductor microring
  resonators.
\newblock {\em IEEE Journal of Selected Topics in Quantum Electronics},
  8(3):705--713, 2002.

\bibitem{carmichael2007statistical}
Howard~J Carmichael.
\newblock {\em Statistical methods in quantum optics 2: Non-classical fields}.
\newblock Springer Science \& Business Media, 2007.

\bibitem{molmer1993monte}
Klaus M{\o}lmer, Yvan Castin, and Jean Dalibard.
\newblock Monte carlo wave-function method in quantum optics.
\newblock {\em Journal of the Optical Society of America B}, 10(3):524--538,
  1993.

\bibitem{garraway1994comparison}
BM~Garraway and PL~Knight.
\newblock Comparison of quantum-state diffusion and quantum-jump simulations of
  two-photon processes in a dissipative environment.
\newblock {\em Physical Review A}, 49(2):1266, 1994.

\bibitem{gerry2023introductory}
Christopher~C Gerry and Peter~L Knight.
\newblock {\em Introductory quantum optics}.
\newblock Cambridge university press, 2023.

\bibitem{bamba2011origin}
Motoaki Bamba, Atac Imamo{\u{g}}lu, Iacopo Carusotto, and Cristiano Ciuti.
\newblock Origin of strong photon antibunching in weakly nonlinear photonic
  molecules.
\newblock {\em Physical Review A}, 83(2):021802, 2011.

\bibitem{liang2019antibunching}
Xinyun Liang, Zhenglu Duan, Qin Guo, Cunjin Liu, Shengguo Guan, and Yi~Ren.
\newblock Antibunching effect of photons in a two-level emitter-cavity system.
\newblock {\em Physical Review A}, 100(6):063834, 2019.

\bibitem{qutip}
J.R. Johansson, P.D. Nation, and Franco Nori.
\newblock Qutip 2: A python framework for the dynamics of open quantum systems.
\newblock {\em Computer Physics Communications}, 184(4):1234–1240, April
  2013.

\bibitem{fox2006quantum}
Anthony~Mark Fox.
\newblock {\em Quantum optics: an introduction}, volume~15.
\newblock Oxford University Press, 2006.

\bibitem{zou1990photon}
XT~Zou and L~Mandel.
\newblock Photon-antibunching and sub-poissonian photon statistics.
\newblock {\em Physical Review A}, 41(1):475, 1990.

\bibitem{gazeau2019coherent}
Jean-Pierre Gazeau.
\newblock Coherent states in quantum optics: an oriented overview.
\newblock In {\em Integrability, Supersymmetry and Coherent States: A Volume in
  Honour of Professor V{\'e}ronique Hussin}, pages 69--101. Springer, 2019.

\bibitem{fabre2020modes}
Claude Fabre and Nicolas Treps.
\newblock Modes and states in quantum optics.
\newblock {\em Reviews of Modern Physics}, 92(3):035005, 2020.

\bibitem{misiaszek2022applications}
Marta Misiaszek.
\newblock {\em Applications of single-photon technology}.
\newblock PhD thesis, 2022.

\bibitem{spirtes2000causation}
Peter Spirtes, Clark~N Glymour, and Richard Scheines.
\newblock {\em Causation, prediction, and search}.
\newblock MIT press, 2000.

\end{thebibliography}

\end{document}